\documentclass[a4paper]{jpconf}
\usepackage{graphicx}
\usepackage{hyperref}
\begin{document}
\title{Improving robustness of jet tagging algorithms with adversarial training: exploring the loss surface}

\author{Annika Stein}

\address{III. Physics Institute A, RWTH Aachen University, Germany}

\ead{annika.stein@cern.ch}

\begin{abstract}
In the field of high-energy physics, deep learning algorithms continue to gain in relevance and provide performance improvements over traditional methods, for example when identifying rare signals or finding complex patterns. From an analyst’s perspective, obtaining highest possible performance is desirable, but recently, some attention has been shifted towards studying robustness of models to investigate how well these perform under slight distortions of input features. Especially for tasks that involve many (low-level) inputs, the application of deep neural networks brings new challenges. In the context of jet flavor tagging, adversarial attacks are used to probe a typical classifier‘s vulnerability and can be understood as a model for systematic uncertainties. A corresponding defense strategy, adversarial training, improves robustness, while maintaining high performance.
Investigating the loss surface corresponding to the inputs and models in question reveals geometric interpretations of robustness, taking correlations into account.

\end{abstract}

\section{Introduction}
With powerful machine learning (especially deep learning) algorithms, new physics analyses have been enabled and established ones report improved results over previous iterations that utilized only cut-based strategies, shallow networks or techniques like BDTs~\cite{MLHEPwhitepaper}. For object identification, which serves as a crucial ingredient to various analyses carried out at experiments at the CERN Large Hadron Collider, it is therefore of prime interest to provide highly-performant algorithms, where many features enter complex architectures to capture as much information as possible, including correlations between observables.

Deep Neural Networks are suited to perform the aforementioned difficult tasks like jet flavour identification, and many low-level features related to the jet constituents enter state-of-the-art taggers~\cite{CMSRun2BTAG,DeepJet,ParticleNet,ParT,ATLASRun2FTAG}. With high performance however comes high reliability on the modeling of the involved input features, especially since supervised machine learning techniques utilize labeled simulated samples~\cite{aisafetyhep}. These likely do not capture all detector effects and can be fairly different for non-identical MC generators, when comparing steps like parton showering and hadronization~\cite{ATLASMCtoMC}. Calibration has therefore always been a necessary step towards improving agreement between the domain on which such algorithms have been trained (simulation), and measured data~\cite{ATLAScalib,CMScalib}. Even after applying a high level of scrutiny and utilizing a set of independent control regions, a certain level of disagreement may remain after calibration, becoming increasingly relevant for analyses where derived scale factors factorize for final states with high (b-/c-tagged) jet multiplicities.
\subsection{Related work}
Reliability on low-level features is a common property in several areas of high-energy physics, and so might be the susceptibility towards slightly distorted features which can lead to drastically reduced performance, better known under the term of adversarial attacks yielding adversarial samples~\cite{aisafetyhep, szegedy2014intriguing,goodfellow2015explaining, paper}. The fundamental principle explored in Ref.~\cite{paper} is that such inherent vulnerability can be turned into robustness, when carefully defending against adversarial attacks via adversarial training. There it has been shown that improving robustness can be achieved without loss of performance~\cite{paper}. The technique which has been used extensively is the Fast Gradient Sign Method (FGSM), a first order attack~\cite{goodfellow2015explaining,paper}. After the proof-of-principle had been introduced for a simple multilayer-perceptron architecture~\cite{paper}, the CMS Collaboration has presented a successful application of adversarial training for the succeeding generation of tagging algorithms, extending the strategy to convolutional (followed by recurrent and dense) layers~\cite{dpnote}. One core finding of Ref.~\cite{dpnote} is the relation between adversarial robustness and agreement between data and simulation, being explicitly prominent for light-flavoured jets. The literature also mentions complimentary perspectives where adversarial training does not capture uncertainties~\cite{Ghosh:2021roe,Butter:2022xyj}. The sentiment that theory-induced or generator-dependent modelings can hardly be handled by adversarial methods is evident~\cite{Butter:2022xyj}, however we intend to focus on those mismodelings which can be mitigated by systematic regularization, which may play the main role in promising control regions shown in Ref.~\cite{dpnote}. Another limitation to consider when applying adversarial training against FGSM attacks is that directions into which inputs are shifted are somewhat predictable, and moreover, always treat all features independently with the respective sign of the gradient, thus eliminating the full correlation between features and leaving only discrete choices~\cite{paper}.
When aiming for robust algorithms that offer not only high performance, but also generalization capabilities, the flatness of the underlying loss surface is scrutinized as a proxy for the aforementioned desired qualities of the model. Several approaches utilize the geometric properties of the loss~\cite{li2018visualizing} as a function of the model parameters (weights and bias terms), but a study with respect to the input distributions has yet to be carried out.
\subsubsection{Adversarial attacks versus systematic uncertainties}
While several restrictions have been imposed to keep the artificial shifts of inputs somewhat realistic with respect to typically observed mismodelings, such adversarial methods are reliant on the network's properties. This marks an unphysical scenario, as neither nature nor simulation of processes could have any knowledge of the machine learning algorithms involved to tag the jets in an event. Thus, judging a network's capability to resist adversarial attacks might be biased towards the defense strategy which explicitly mitigates the impact of specific attacks. It is unrealistic that any mismodeling in simulation would shift inputs exclusively in the worst case direction pointing to steepest increase of the loss function~\cite{aisafetyhep, paper}. For physics analysis, it is not of primary relevance to utilize algorithms which are robust against adversarial attacks, but which allow generalization from simulation to data and offer robustness towards systematic uncertainties. Therefore, the two trainings studied in Ref.~\cite{paper} (nominal and adversarial) are compared not only with nominal and adversarial inputs, but also when being exposed to systematically distorted inputs which point either in upwards or downwards direction~\cite{paper}.
In both cases, up- or downwards variation, adversarial training performs better on distorted inputs than nominal training on same distorted samples~\cite{paper}. Similar conclusions can be drawn when exchanging the systematic variations with random smearing / Gaussian noise~\cite{paper}. In this paper we intend to augment the findings by investigating the underlying loss function in the input feature space to propose a modified training strategy which can improve the algorithm's resilience.
\section{Properties of loss manifolds for a jet tagging algorithm trained on nominal or adversarial samples}
\begin{figure}[h]
\begin{center}
\includegraphics[width=0.8\textwidth]{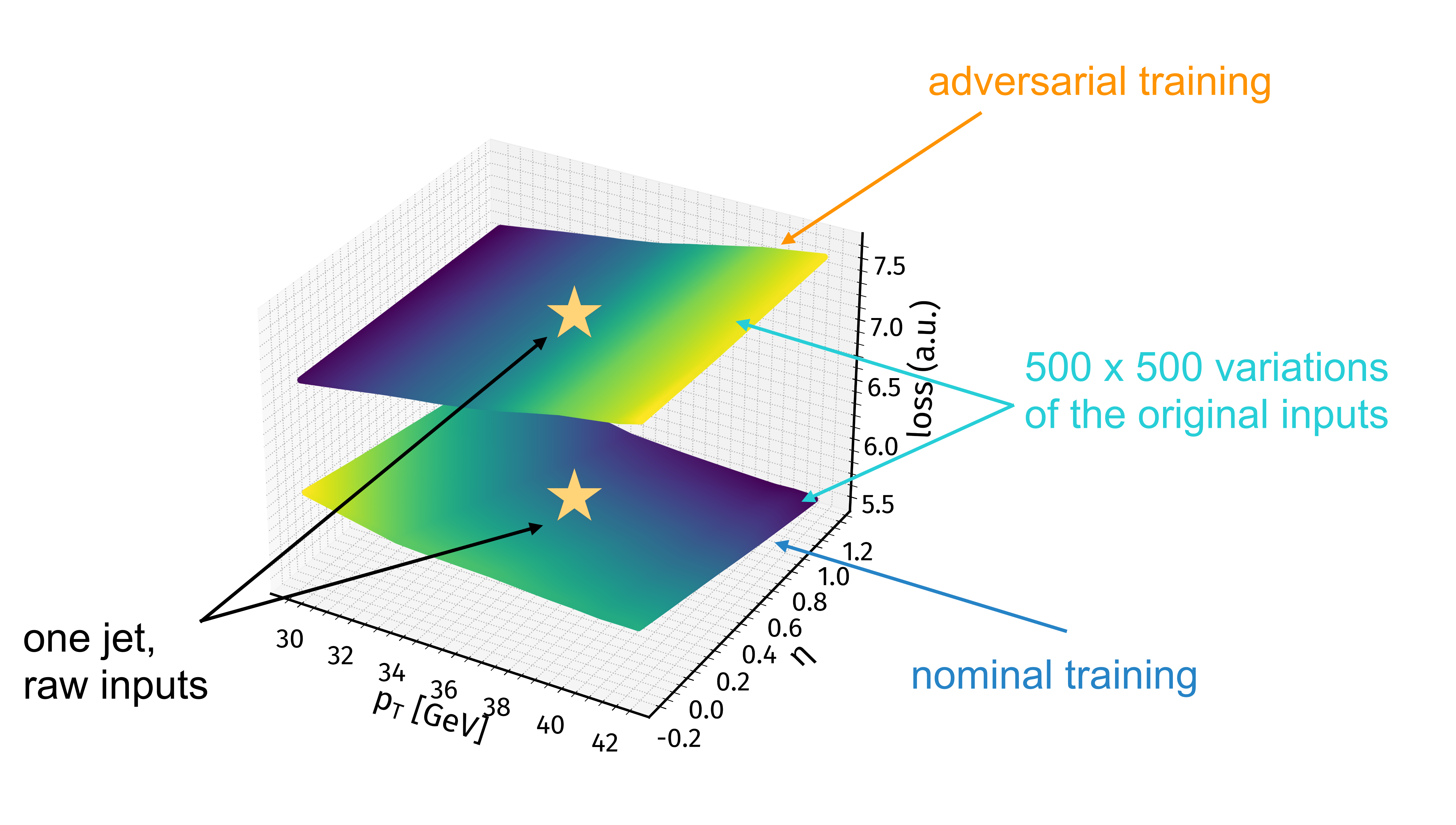}
\end{center}
\caption{\label{fig:loss_with_desc}Different geometries of loss manifolds for nominal (bottom) and adversarial (top) training.}
\end{figure}

The assumption of different geometry of loss manifolds has been motivated by observations made when looking at the impact of adversarial attacks split by flavour, where adversarial training behaves somewhat symmetrically, but adversarial attacks performed for nominal training push inputs preferably into specific directions to invert expected physics~\cite{paper}. While the illustrations presented in Ref.~\cite{paper} give a hint on how the loss surfaces of different training strategies could look like, it has been an open question to perform realistic scans of such surfaces. First results of such a visualization of geometry with respect to input variations are presented in Fig.~\ref{fig:loss_with_desc}. The construction is obtained by first selecting a random jet drawn from a sample which has not been used for training or validation. Focusing on two observables (for visualization purposes, using well-understood global jet features), a grid of $500\times 500$ variations is generated, using a uniform and symmetric binning around the original nominal features. Taking the full distribution of the respective feature into account, the spanned range corresponds to $\pm 0.5\sigma$, ensured by only working in the input feature space after standardization. While the target remains unchanged, both the nominal and adversarial training are reevaluated on the resulting $250000$ samples, and the resulting loss is recalculated.
Moving a jet's pseudorapidity without changing transverse momentum or other properties will not affect the respective network prediction error, or loss, for adversarial training. Nominal training on the other hand is not agnostic to changes in any of the two variables shown. While nominal training offers in general a lower network prediction error, adversarial training offers a flatter manifold with a certain level of invariance with respect to distortions of specific features.
\begin{figure}
\includegraphics[width=18pc]{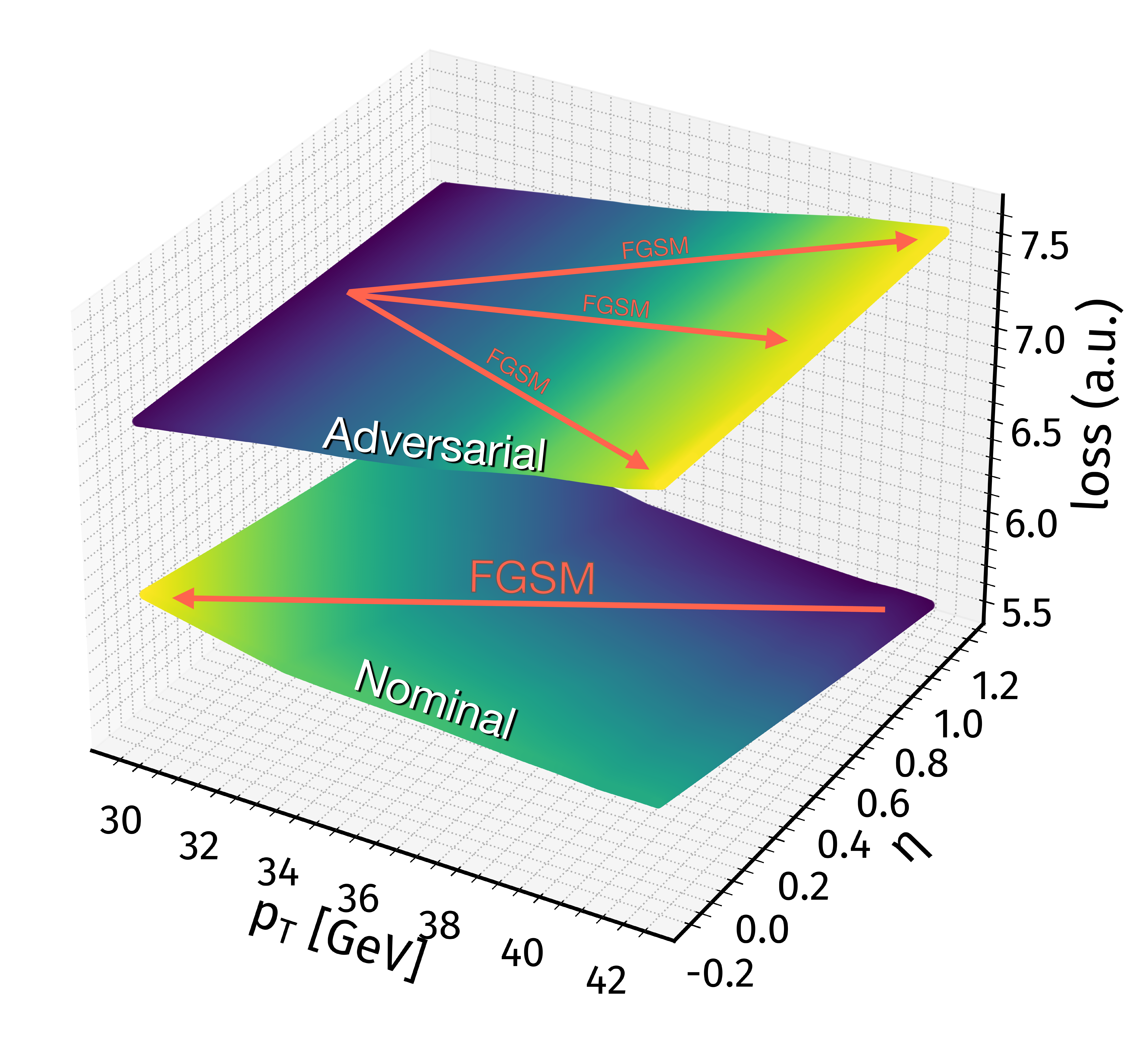}\hspace{2pc}%
\begin{minipage}[b]{14pc}\caption{\label{fig:loss_with_directions}Possible directions of adversarial attacks for different models. Starting from kinematic quantities which yield small loss, multiple arrows can be found for an FGSM attack imposed for adversarial training, while only one such arrow is constructed for nominal training.}
\end{minipage}
\end{figure}
Figure~\ref{fig:loss_with_directions} reveals how for nominal training, adversarial attacks would find a clear direction, while for adversarial training, due to the invariance or symmetry with respect to pseudorapidity, multiple directions are possible to increase the loss. Despite this finding, only one specific direction will be chosen by the attack as a result of the inherent operation of taking the sign of the gradient, although other directions would essentially lead to the same effect. 
\section{Discussion}
This observation is a key element to understand why adversarial training may be preferred in settings with potentially distorted inputs (due to experimental effects, precision and resolution limitations) or other systematic differences between the domain on which the identification algorithms have been trained (simulation) and the domain built from actually recorded detector data. This can be interpreted as a sign of regularization induced by adversarial training.
\subsection{Using properties of the loss manifold during training}
Having probed the loss manifold on a macro-scale for two features which may not offer highest discriminating power (and which have been reweighted to a common target distribution to ensure bias-free predictions)~\cite{paper}, we propose to explore this technique more systematically and potentially incorporate this into the training itself. Showing the loss surface as a function of two input features is a simplification which allows us to investigate the geometry graphically. To overcome this limitation, the loss manifold needs to be constructed in several more dimensions in the feature space. Then, measuring flatness around the original inputs can be introduced as an independent cross-check during training to probe and improve robustness. We can construct a summary quantity as an additional term in the loss function, for example capturing the maximally observed relative impact on the calculated cross-entropy loss when moving inputs in the allowed $B_{\frac{\sigma}{2}}$-ball. This can be weighted by a hyperparameter to control how much focus is given to regularization, compared to plain performance metrics, and training would then follow this modified loss function during backpropagation to update the model parameters.
\subsection{Building other attacks which preserve directionality of the gradient of the loss function}
From the observed loss surfaces it seems sufficient to continue focusing on first order attacks, although taking the sign of the gradient (FGSM) might be too inefficient when actual directions of gradients and relative contributions of features are to be taken into account. Using the $p$-norm of gradients where e.g. $p=2$ instead, the individual input feature's contribution can be maintained quantitatively. The resulting distortion vector can be scaled by the inverse of the aforementioned norm to allow comparisons across different jet samples, while at the same time yielding small disturbances only. This leads to an attack which is not easy to predict, both for the direction of the shift, as well as the magnitude per feature, unlike for FGSM, where only $\pm\epsilon$ shifts are possible. Introducing the modified attack instead will include correlations between features, a shortcoming of the FGSM attack typically mentioned in the context of HEP. In an adversarial training against this new attack, we would not need large distortions, resulting distorted jet samples will not be easy to detect in validation methods (such as one-dimensional histograms).
\section{Conclusion}
In this paper, we presented a study of the loss manifold with respect to input features of a typical jet tagging algorithm, when trained on nominal or adversarial samples. Differences with respect to flatness and thus invariance to small distortions are observed, explaining and confirming previously explored differences in robustness and generalization. With such loss surfaces at hand, we proposed modified training strategies to explicitly use that newly gained knowledge of the network's properties directly during backpropagation. Putting more focus on regularization and correlations, the proposed methods can bridge the gap between machine learning-theoretical studies and their application for object identification in particle physics, where the physical behaviour of observables shall be maintained.

\ack
Simulations were performed with computing resources granted by RWTH Aachen University under project \texttt{rwth1244}. This work has received support by the Deutsche Forschungsgemeinschaft (DFG, German Research Foundation, projects SCHM 2796/5 and GRK 2497), and the Bundesministerium für Bildung und Forschung (BMBF, Project 05H2021). 
\section*{References}
\bibliographystyle{iopart-num}
\bibliography{iopart-num}

\end{document}